\documentclass[10pt]{amsart}
\usepackage{amsmath,amssymb}
\usepackage[dvips]{epsfig}

\newtheorem{theorem}{Theorem}[section]
\newtheorem{lemma}[theorem]{Lemma}

\allowdisplaybreaks \numberwithin{equation}{section}

\def\Tr{{\rm Tr}}

\def\rank{\mathop{\rm rank}\nolimits}
\newcommand{\PP}{{\mathbb P}}
\def\Im{\mathop{\rm Im}\nolimits}
\def\Re{\mathop{\rm Re}\nolimits}

\def\Jac{\mathop{\rm Jac}\nolimits}

\def\diag{\mathop{\rm Diag}\nolimits}

\def\Hilb{\mathop{\rm Hilb}\nolimits}

\def\Cof{\mathop{\rm Cof}\nolimits}

\begin{document}

\title[Cyclic Monopoles]
{Cyclic Monopoles, Affine Toda and Spectral Curves}
\author{H.W. Braden}
\address{School of Mathematics, Edinburgh University, Edinburgh.}
\email{hwb@ed.ac.uk}

\begin{abstract}
We show that any cyclically symmetric monopole is gauge equivalent
to Nahm data given by Sutcliffe's ansatz, and so obtained from the
affine Toda equations. Further the direction (the Ercolani-Sinha
vector) and base point of the linearising flow in the Jacobian of
the spectral curve associated to the Nahm equations arise as
pull-backs of Toda data. A theorem of Accola and Fay then means that
the theta-functions arising in the solution of the monopole problem
reduce to the theta-functions of Toda.

\end{abstract}

 \maketitle

\tableofcontents

\section{Introduction}

Magnetic monopoles, the topological soliton solutions of
Yang-Mills-Higgs gauge theories in three space dimensions with
particle-like properties, have been the subject of considerable
interest over the years. BPS monopoles, arising from a limit in
which the the Higgs potential is removed but a remnant of this
remains in the boundary conditions, satisfy the first order
Bogomolny equation
$$B_i=\frac{1}{2}\sum_{j,k=1}\sp3\epsilon_{ijk}F\sp{jk}=D_i\Phi$$
and have merited particular attention (see \cite{ms04} for a recent
review). This focus is in part due to the ubiquity of the Bogomolny
equation.  Here $F_{ij}$ is the field strength associated to a gauge
field $A$, and $\Phi$ is the Higgs field. We shall focus on the case
when the gauge group is $SU(2)$. The Bogomolny equation may be
viewed as a dimensional reduction of the four dimensional self-dual
equations upon setting all functions independent of $x_4$ and
identifying $\Phi=A_4$; they are also encountered in supersymmetric
theories when requiring certain field configurations to preserve
some fraction of supersymmetry. The study of BPS monopoles is
intimately connected with integrable systems. Nahm gave a transform
of the ADHM instanton construction to produce BPS monopoles
\cite{nahm82} and the resulting Nahm's equations have Lax form with
corresponding spectral curve $\hat{\mathcal{C}}$. This curve,
investigated by Corrigan and Goddard \cite{cg81}, was given a
twistorial description by Hitchin \cite{hitchin82} where the same
curve lies in mini-twistor space, $\hat{\mathcal{C}}\subset$
T$\mathbb{P}\sp1$. Just as Ward's twistor transform relates
instanton solutions on $\mathbb{R}\sp4$ to certain holomorphic
vector bundles over the twistor space $\mathbb{CP}\sp3$, Hitchin
showed that the dimensional reduction leading to BPS monopoles could
be made at the twistor level as well and was able to prove that all
monopoles could be obtained by this approach \cite{hitchin83}
provided the curve $\hat{\mathcal{C}}$ was subject to certain
nonsingularity conditions. Bringing methods from integrable systems
to bear upon the construction of solutions to Nahm's equations for
the gauge group $SU(2)$ Ercolani and Sinha \cite{es89} later showed
how one could solve (a gauge transform of) the Nahm equations in
terms of a Baker-Akhiezer function for the curve
$\hat{\mathcal{C}}$.

Although many general results have now been obtained few explicit
solutions are known. This is for two reasons, each coming from a
transcendental constraint on the curve $\hat{\mathcal{C}}$. The
first is that the curve $\hat{\mathcal{C}}$ is subject to a set of
constraints whereby  the periods of a meromorphic differential on
the curve are specified. This type of constraint arises in many
other settings as well, for example when specifying the filling
fractions of a curve in the AdS/CFT correspondence. Such constraints
are transcendental in nature and until quite recently these had only
been solved in the case of elliptic curves (which correspond to
charge $2$ monopoles). In \cite{bren06,bren07} they were solved for
a class of charge $3$ monopoles using number theoretic results of
Ramanujan. The second type of constraint is that the linear flow on
the Jacobian of $\hat{\mathcal{C}}$ corresponding to the integrable
motion only intersects the theta divisor in a prescribed manner. In
the monopole setting this means the Nahm data will yield regular
monopole solutions but a similar constraint also appears in other
applications of integrable systems. In Hitchin's approach (reviewed
below) this may be expressed as the vanishing of a real one
parameter family of cohomologies of certain line bundles,
$H^0(\hat{\mathcal{C}},L^{\lambda}(n-2))=0$ for $\lambda\in(0,2)$.
Viewing the line bundles as points on the Jacobian this is
equivalent to a real line segment not intersecting the theta divisor
$\Theta$ of the curve. Indeed there are sections for $\lambda=0,2$
and the flow is periodic (mod $2$) in $\lambda$ and so we are
interested in the number of times a real line intersects $\Theta$.
While techniques exist that count the number of intersections of a
complex line with the theta divisor we are unaware of anything
comparable in the real setting and again solutions have only been
found for particular curves \cite{bren09}. Thus the application of
integrable systems techniques to the construction of monopoles and
(indeed more generally) encounters two types of problem that each
merit further study.

The present paper will use symmetry to reduce these problems to
ones more manageable. Long ago monopoles of charge $n$ with cyclic
symmetry $\texttt{C}_n$ were shown to exist \cite{or82} and more
recently such monopoles were reconsidered \cite{hmm95} from a
variety of perspectives. The latter work indeed considered the
case of monopoles with more general Platonic symmetries and for
the case of tetrahedral, octahedral and icosahedral symmetry
(where such monopoles exist) the curves were reduced to elliptic
curves. (See \cite{hs96a, hs96b, hs97} for development of this
work.) Our first result is to strengthen work of Sutcliffe
\cite{sut96}. Motivated by Seiberg-Witten theory Sutcliffe gave an
ansatz for $\texttt{C}_n$ symmetric monopoles in terms of $su(n)$
affine Toda theory. The spectral curve $\hat{\mathcal{C}}$ of a
$\texttt{C}_n$ symmetric monopole yields an $n$-fold unbranched
cover of the hyperelliptic spectral curve $\mathcal{C}$ of the
affine Toda theory, a spectral curve that arises in Seiberg-Witten
theory describing the pure gauge $\mathcal{N}=2$ supersymmetric
$su(n)$ gauge theory. (We shall recall some properties of the Nahm
construction and this relation between curves in section 2.)
Sutcliffe's ansatz (section 3) shows how solutions to the affine
Toda equations yield cyclically symmetric monopoles. Our first
result proves that any cyclically symmetric monopole is gauge
equivalent to Nahm data given by Sutcliffe's ansatz, and so
obtained from the affine Toda equations. We mention that Hitchin
in an unpublished note had, prior to Sutcliffe, observed that
cyclic charge 3 monopoles were equivalent to solutions of the
affine Toda equations. The remainder of this paper shows that the
relation between the Nahm data and the affine Toda system is much
closer than simply that they yield the same equations of motion.
The solution of an integrable system is typically expressed in
terms of the straight line motion on the Jacobian of the system's
spectral curve. Such a line is determined both by its direction
and a point on the Jacobian. We shall show that both the direction
(given by the Ercolani-Sinha vector, section 4) and point relevant
for monopole solutions (section 5) are obtained as pull-backs of
Toda data. This connection is remarkable and ties the geometry
together in a very tight manner. Section 6 recalls a theorem of
Accola and Fay that holds in precisely this setting, showing how
the theta-functional solutions of the monopole reduce to precisely
the theta-functional solutions of Toda. At this stage we have
reduced the problem of constructing cyclically symmetric monopoles
to one of determining hyperelliptic curves that satisfy the
transcendental constraints described above. Though more manageable
the problems are still formidable and a construction in the charge
$3$ setting will be described elsewhere \cite{bde10}. We conclude
with a discussion.

\section{Monopoles}
We shall briefly recall the salient features for constructing
$su(2)$ monopoles of charge $n$. We begin with Nahm's construction
\cite{nahm82}. In generalizing the ADHM construction of instantons
Nahm established an equivalence between nonsingular monopoles and
what is now referred to as Nahm data: three $n\times n$ matrices
$T_i(s)$ with $s\in[0,2]$ satisfying
\begin{itemize}
\item[\bf{N1}] Nahm's equation
\begin{equation}
\frac{dT_i}{ds}=\frac{1}{2}\sum_{j,k=1}^3\epsilon_{ijk}[T_j,T_k],
\end{equation}

\item[\bf{N2}] $T_i(s)$ is regular for $s\in(0,2)$ and has simple
poles at $s=0$ and $s=2$, the residues of which form an irreducible
$n$-dimensional representation of $su(2)$,

\item[\bf{N3}]
$\displaystyle{T_i(s)=-T_i^\dagger(s),\qquad T_i(s)=T_i^t(2-s). }$
\end{itemize}
Upon defining
\begin{align*}
A(\zeta)&=T_1+iT_2-2iT_3\zeta+(T_1-iT_2)\zeta\sp2\\
M(\zeta)&=-iT_3+(T_1-iT_2)\zeta \intertext{we find that Nahm's
equation is equivalent to the Lax equation}
\frac{dT_i}{ds}&=\frac{1}{2}\sum_{j,k=1}^3\epsilon_{ijk}[T_j,T_k]
\Longleftrightarrow
    [\dfrac{d}{ds}+M,A]=0.
\end{align*}
Here $\zeta$ is a spectral parameter. Following from the Lax
equation we have the invariance of the spectral curve
\begin{equation}\label{spectcurve}
\hat{\mathcal{C}}:\ 0=P(\eta,\zeta):=\det(\eta 1_n+A(\zeta))
\end{equation}
where
\begin{equation}\label{spectcurveP}
P(\eta,\zeta)=\eta\sp{n}+a_1(\zeta)\eta\sp{n-1}+\ldots+a_n(\zeta),\qquad
\deg a_r(\zeta)\le2r.\end{equation} As with any spectral curve
presented in the form (\ref{spectcurve}) one should always ask
where $\hat{\mathcal{C}}$ lies. Typically the spectral curve lies
in a surface, $\hat{\mathcal{C}}\subset\mathcal{S}$, and
properties of the surface are closely allied with the integrable
system encoded by the Lax equation. (For example, for suitable
surfaces, the separation of variables may be described by
$\Hilb\sp{[N]}(\mathcal{S})$, the Hilbert scheme of points on
$\mathcal{S}$.) For the case at hand
\begin{equation*}
{\hat{\mathcal{C}}}
\subset
    T\mathbb{P}\sp1:=\mathcal{S},\qquad (\eta,\zeta)\rightarrow
    \eta\frac{d}{d\zeta}\in T\mathbb{P}\sp1,
\end{equation*}
and monopoles admit a minitwistor description: the curve
$\hat{\mathcal{C}}$ corresponds to those lines in $\mathbb{R}\sp3$
which admit normalizable solutions of an appropriate scattering
problem in both directions \cite{hitchin82, hitchin83}. This latter
description makes clear that $\hat{\mathcal{C}}$ comes equipped with
an antiholomorphic involution or real structure coming from the
reversal of orientation of lines
    $$(\eta,\zeta)\rightarrow(-\bar\eta/{\bar\zeta}\sp2,-1/{\bar\zeta}).$$
This means the coefficients of (\ref{spectcurveP}) are such that
\begin{equation}\label{reality}
{a_r(\zeta)=(-1)\sp{r}\zeta\sp{2r}\overline{a_r(-1/{\overline\zeta}\,)}}
\end{equation}
and so each may be expressed in terms of $2r+1$ (real) parameters
$$ a_r(\zeta)= \chi_r\, \left[\prod_{l=1}\sp{r}\left(
\frac{\overline{\alpha}_{r,l}}{\alpha_{r,l}}\right)\sp{1/2}\right]
\prod_{k=1}\sp{r}(\zeta-\alpha_{r,k})(\zeta+\frac{1}{\overline{\alpha}_{r,k}}),\qquad
\alpha_{r,k}\in \mathbb{C},\ \chi_r\in\mathbb{R}. $$ We remark that a
real structure constrains the form of the period matrix of a curve
and that while in general there may be between $0$ and $\hat g+1$
ovals of fixed points of an antiholomorphic involution ($\hat g$
being the genus of $\hat{\mathcal{C}}$) for the case at hand there
are no fixed points. For the monopole spectral curve
(\ref{spectcurveP}) we have (generically) $\hat g=(n-1)\sp2$.

Although in many situations the solution of the integrable system
encoded by a Lax pair (with spectral parameter) only depends on
intrinsic properties of the spectral curve the monopole physical
setting means that extrinsic properties of our curve in
$T\mathbb{P}\sp1$ are relevant here. Spatial symmetries act on the
monopole spectral curve via fractional linear transformations.
Although a general M\"obius transformation does not change the
period matrix of a curve $\hat{\mathcal{C}}$ only the subgroup
$PSU(2)< PSL(2,\mathbb{C})$ preserves the reality properties
necessary for a monopole spectral curve. These reality conditions
are an extrinsic feature of the curve (encoding the space-time
aspect of the problem) whereas the intrinsic properties of the curve
are invariant under birational transformations or the full M\"obius
group. Such extrinsic aspects are not a part of the usual integrable
system story. Thus $SO(3)$ spatial rotations induce an action on
$T\PP\sp1$ via $PSU(2)$: if $\begin{pmatrix} p&q\\-\bar q&\bar
p\end{pmatrix}\in PSU(2)$, ($ |p|\sp2+|q|\sp2=1$) then
\begin{equation}\label{pstransf}
\zeta\rightarrow\tilde\zeta:=\dfrac{\bar p\, \zeta-\bar q}{q\,
\zeta+p}, \qquad \eta\rightarrow\tilde\eta:= \dfrac{\eta}{(q\,
\zeta+p)\sp2}
\end{equation}
corresponds to a rotation by $\theta$ around ${\mathbf n}\in S\sp2$
where $$ n_1\sin{(\theta/2)}=\Im q,\ n_2\sin{(\theta/2)}=-\Re q,\
 n_3\sin{(\theta/2)}=\Im p,\
\cos{(\theta/2)}=-\Re p.
$$
This $SO(3)$ action commutes with the standard real structure on
$T\PP\sp1$. The action on the spectral curve may be expressed as
\begin{equation}
P(\tilde\eta,\tilde\zeta)=\dfrac{{\tilde P}(\eta,\zeta)}{(q\,
\zeta+p)\sp{2n}}, \qquad {\tilde
P}(\eta,\zeta)=\eta\sp{n}+\sum_{r=1}\sp{n}\eta\sp{n-r}\,{\tilde
a}_r(\zeta),
\end{equation}
where in terms of the parameterization above
$$
a_r(\zeta)\rightarrow \frac{{\tilde a}_r(\zeta)}{(q\,
\zeta+p)\sp{2r}}\equiv \frac{{\tilde\chi}_r}{(q\, \zeta+p)\sp{2r}}
\left[\prod_{l=1}\sp{r}\left(
\frac{\overline{\tilde\alpha}_l}{\tilde\alpha_l}\right)\sp{1/2}\right]
\prod_{k=1}\sp{r}(\zeta-{\tilde\alpha}_k)(\zeta+\frac{1}{\overline{\tilde\alpha}_k})
$$
with
$$\alpha_k\rightarrow {\tilde\alpha}_k \equiv\frac{p\alpha_k+{\bar
q}}{{\bar p}-\alpha_k q},\qquad \chi_r \rightarrow
{\tilde\chi}_r\equiv \chi_r \prod_{k=1}\sp{r}\left[\frac{({\bar
p}-\alpha_k q)({ p}-{\bar\alpha_k}{\bar q})({\bar\alpha_k}{\bar
p}+q)(\alpha_k{ p}+ {\bar
q})}{{\alpha_k}{\bar\alpha_k}}\right]\sp{1/2}.$$ In particular the
form of the curve does not change under a rotation: that is, if
$a_r=0$ then so also ${\tilde a}_r=0$.

Hitchin, Manton and Murray \cite{hmm95} showed how curves
invariant under finite subgroups of $SO(3)$ or their binary covers
yield symmetric monopoles. Suppose we have a symmetry; the
spectral curve $0=P(\eta,\zeta)$ is transformed to the same curve,
$0=P(\tilde\eta,\tilde\zeta)= {{\tilde P}(\eta,\zeta)}/{(q\,
\zeta+p)\sp{2n}}$. Then $P(\eta,\zeta)={\tilde P}(\eta,\zeta)$, or
equivalently $a_r(\zeta)={\tilde a}_r(\zeta)$. Relevant for us is
the example of cyclically symmetric monopoles. Let
$\omega=\exp(2\pi i/n)$. A rotation of order $n$ is then given by
$\bar p=\omega\sp{1/2}$, $q=0$ which yields
$$\phi:\ (\eta,\zeta)\rightarrow (\omega\eta,\omega\zeta).$$
Correspondingly $\eta\sp{i}\zeta\sp{j}$ is invariant for $i+j\equiv
0 \mod n$ and the spectral curve
$$\eta\sp{n}+a_1 \eta\sp{n-1}\zeta+a_2
\eta\sp{n-2}\zeta\sp2+\ldots+a_n\zeta\sp{n}+\beta\zeta\sp{2n}+\gamma=0$$
is invariant under the cyclic group $\texttt{C}_n$ generated by this
rotation. Imposing the reality conditions (\ref{reality}) and
centering the monopole (setting $a_1=0$) then gives us the spectral
curve in the form \begin{equation}\label{cycliccurve}\eta\sp{n}+a_2
\eta\sp{n-2}\zeta\sp2+\ldots+a_n\zeta\sp{n}+\beta\zeta\sp{2n}+(-1)\sp{n}\bar\beta=0,\qquad
a_i\in\mathbf{R} \end{equation} and by an overall rotation we may
choose $\beta$ real.

Now the $\texttt{C}_n$-invariant curve $\hat{\mathcal{C}}$
(\ref{cycliccurve}) of genus $\hat g=(n-1)^2$ is an $n$-fold
unbranched cover of a genus $g=n-1$ curve $\mathcal{C}$. The
Riemann-Hurwitz theorem yields the relation $\hat g =n(g-1)+1$.
Introduce the rational invariants $x=\eta/\zeta$, $
\nu=\zeta\sp{n}\beta$, then
$$x\sp{n}+a_2 x\sp{n-2}+\ldots
+a_n+\nu+\frac{(-1)\sp{n}|\beta|\sp2}{\nu}=0$$ and upon setting
$y=\nu-{(-1)\sp{n}|\beta|\sp2}/{\nu}$ we obtain the curve
\begin{equation}\label{Todacurve}
y^2=(x\sp{n}+a_2 x\sp{n-2}+\ldots +a_n)\sp2-4(-1)\sp{n}|\beta|\sp2.
\end{equation}
This curve is the spectral curve of $su(n)$ affine Toda theory in
standard hyperelliptic form.

For future reference we note that the $n$-points $\hat\infty_j$
above the point $\zeta=\infty$ project to one of the infinite
points, $\infty_+$, of the curve (\ref{Todacurve}), while the
$n$-points above the point $\zeta=0$ project to the other infinite
point. At $\hat\infty_j$ we have ${\eta}/{\zeta}\sim\rho_j \zeta$ as
$ \zeta\sim\hat{\infty}_j$, with $\rho_j=\beta\sp{1/n}\exp(2\pi
i[j+1/2]/n)$.

\noindent{{\bf{{The $n=2$ example:}}} The reality conditions for
$n=2$ and $a_2(\zeta)=\beta \zeta^4+\gamma\zeta^2+\delta$ means
that $\delta=\bar\beta$ and $\gamma=\bar\gamma$ and
(\ref{cycliccurve}) becomes
$$\eta^2+\beta\zeta^4+\gamma\zeta^2+\bar\beta=0.$$ This is an
elliptic curve. If $\beta=|\beta|e\sp{2i\theta}$ let $U=\zeta
e\sp{i\theta}$ and $V=i\eta e\sp{i\theta}/|\beta|\sp{1/2}$ and
this may be rewritten as
\begin{equation}\label{examplen2a}
V^2=U^4+t\,U^2+1, \qquad
t=\gamma/|\beta|.
\end{equation}
For irreducibility $t\ne2$. Now the curve (\ref{Todacurve})
becomes (with $Y=y/\sqrt{\gamma^2-4|\beta|\sp2}$)
\begin{equation}\label{examplen2b}
Y^2=x^4+t'\,x^2+1, \qquad t'=\frac{2t}{\sqrt{t^2-4}}.
\end{equation}
These two curves (\ref{examplen2a}, \ref{examplen2b}) are
$2$-isogenous: if we quotient the former curve under the
involution $(U,V)\rightarrow(-U,-V)$ we obtain the latter.

\section{The Sutcliffe Ansatz}

Some years ago Sutcliffe \cite{sut96} introduced the following
ansatz for cyclically symmetric monopoles. Let
\begin{align}
T_1+iT_2&=\begin{pmatrix} 0&e\sp{(q_1-q_2)/2}&0&\ldots&0\\
0&0&e\sp{(q_2-q_3)/2}&\ldots&0\\
\vdots&&&\ddots&\vdots\\
0&0&0&\ldots&e\sp{(q_{n-1}-q_n)/2}\\
e\sp{(q_n-q_1)/2}&0&0&\ldots&0
\end{pmatrix}\\
T_1-iT_2&=-\begin{pmatrix}0&0&\ldots&0&e\sp{(q_n-q_1)/2}\\
e\sp{(q_1-q_2)/2}&0&\ldots&0&0\\
0&e\sp{(q_2-q_3)/2}&\ldots&0&0\\
\vdots&&\ddots&&\vdots\\
0&0&\ldots&e\sp{(q_{n-1}-q_n)/2}&0
\end{pmatrix}\\
T_3&=-\frac{i}{2}\begin{pmatrix} p_1&0&\ldots&0\\
0&p_2&\ldots&0\\
\vdots&&\ddots&\vdots\\
0&0&\ldots&p_n
\end{pmatrix}
\end{align}
where $p_i$, $q_i$ are real. Then $T_i(s)=-T_i^\dagger(s)$ and
Nahm's equations yield
\begin{align*}
&\frac{d}{ds}\left(T_1+iT_2\right)=i[T_3,T_1+iT_2]&\Rightarrow&
\begin{cases}p_1-p_2=\dot q_1-\dot q_2,\\
\qquad\vdots\\ p_n-p_1=\dot q_n-\dot q_1.
\end{cases}\\
&\frac{d}{ds}T_3=[T_1,T_2]=\frac{i}2[T_1+iT_2,T_1-iT_2]&\Rightarrow&
\begin{cases}
\dot p_1=-e\sp{q_1- q_2}+e\sp{q_n-q_1},\\
\qquad\vdots \\ \dot p_n=-e\sp{q_n- q_1}+e\sp{q_{n-1}-q_n}.
\end{cases}
\end{align*}
These equations then follow from the equations of motion of the
affine Toda Hamiltonian
\begin{equation}\label{affToda}
H=\frac12\left(p_1\sp2+\ldots+p_n\sp2\right)-\left[e\sp{q_1-
q_2}+e\sp{q_2- q_3}+\ldots+e\sp{q_{n}- q_1} \right] .
\end{equation}
Sutcliffe's observation is that particular solutions of these
equations will then yield cyclically invariant monopoles. In fact
the monopole Lax operator $A(\zeta)$ here is essentially the usual
Toda Lax operator and $$\frac12\Tr A(\zeta)\sp2=\zeta\sp2\, H.$$
The spectral curve of the affine Toda system is then
(\ref{Todacurve}) upon restricting the center of mass motion
$\sum_i p_i=0=\sum_i q_i$. The constant $\beta$ may be related to
the coefficient of the scaling element when the Toda equations are
expressed in terms of the affine algebra $\widehat{\text{sl}}_n$.

In fact we may strengthen Sutcliffe's ansatz substantially. At this
stage we only have that solutions of the Toda equations will yield
some solutions of the Nahm equations with cyclic symmetry. First we
will show that any $\texttt{C}_n$ invariant solution of Nahm's
equations (for charge $n$ $su(2)$ monopoles) are given by solutions
of the affine Toda equations. Then we will very concretely relate
the solutions.

We have that $G\subset SO(3)$ acts on triples
${\bf{t}}=(T_1,T_2,T_3)\in\mathbb{R}\sp3\otimes SL(n,\mathbb{C})$
via the natural action on $\mathbb{R}\sp3$ and conjugation on
$SL(n,\mathbb{C})$. This natural action may be identified with the
$SU(2)$ action on $\mathcal{O}(2)$ given above. If $g'\in SO(3)$ and
$g=\rho(g')$ is its image in $SL(n,\mathbb{C})$ then we have
\begin{align*}g'\circ&\left[
\eta+(T_1+iT_2)-2iT_3\zeta+(T_1-iT_2)\zeta\sp2\right]\\&=
\omega\left[
\eta+\omega\sp{-1}g(T_1+iT_2)g\sp{-1}-2igT_3g\sp{-1}\zeta+ \omega
g(T_1-iT_2)g\sp{-1}\zeta\sp2\right] .\end{align*} Thus invariance of
the spectral curve gives
\begin{align*}
g(T_1+iT_2)g\sp{-1}&=\omega(T_1+iT_2), \\ gT_3g\sp{-1}&=T_3, \\
g(T_1-iT_2)g\sp{-1}&=\omega\sp{-1}(T_1-iT_2).
\end{align*}
Now Hitchin, Manton and Murray \cite{hmm95} have described how the
$SO(3)$ action on $SL(n,\mathbb{C})$ decomposes as the direct sum
$\underline{2n-1}\oplus
\underline{2n-3}\oplus\ldots\oplus\underline{5}\oplus\underline{3}$
where $\underline{2k-1}$ denotes the $SO(3)$ irreducible
representation of dimension $2k-1$. We may identify $SO(3)$ and
its image in $SL(n,\mathbb{C})$ and because this decomposition has
$\rank SL(n,\mathbb{C})=n-1$ summands then, by a theorem of
Kostant \cite{kostant}, the Lie algebra of this $SO(3)$ is a
principal three-dimensional subalgebra. By conjugation we may
express our generator $g'$ of
$\texttt{C}_n$ as $g'=\exp\left[\dfrac{2\pi}{n}\begin{pmatrix}0&1&0\\
-1&0&0\\0&0&0\end{pmatrix}\right]$ and then
$g=\rho(g')=\exp\left[\frac{2\pi}{n}H\right]$ where $H$ is
semi-simple and the generator of the principal three-dimensional
algebra's Cartan subalgebra. Kostant described the action of such
elements on arbitrary semi-simple Lie algebras and their roots.
For the case at hand we have that $g$ is equivalent to
$\diag(\omega\sp{n-1},\ldots,\omega,1)$ and that
$$g E_{ij}g\sp{-1}=\omega\sp{j-i}\,E_{ij}.$$
Therefore at this stage we know that for a cyclically invariant
monopole we may write
$$T_1+iT_2=\sum_{\alpha\in\hat \Delta}e\sp{(\alpha,\tilde
q)/2}\,E_\alpha,\qquad T_3=-\frac{i}2\sum_j \tilde p_j\,H_j$$
where in principle $\tilde q_i$, $\tilde p_i\in\mathbb{C}$, and
$\alpha\in\hat \Delta$ are the simple roots together with minus
the highest root. (The sum over $H_i$ may be taken as either the
Cartan subalgebra of $SL(n,\mathbb{C})$ or, by reinstating the
center of mass, the Cartan subalgebra of $GL(n,\mathbb{C})$.) The
Sutcliffe ansatz follows if the $\tilde q_i$ and $\tilde p_i$ may
be chosen real. Now by an $SU(n)$ transformation
$\diag(e\sp{i\theta_1},e\sp{i\theta_2},\ldots, e\sp{i\theta_n})$
(where $\sum_i\theta_i=0$) together with an overall $SO(3)$
rotation the reality of $\tilde q_i$ may be achieved. The reality
of $\tilde p_i$ follows upon imposing $T_i(s)=-T_i\sp\dagger(s)$
which also fixes $T_1-iT_2$. At this stage we have established the
following.
\begin{theorem} Any cyclically symmetric monopole is gauge equivalent
to  Nahm data given by Sutcliffe's ansatz, and so obtained from the
affine Toda equations.
\end{theorem}

\section{Flows and Solutions}
The relation between the Nahm data and the affine Toda system is
much closer than simply that they yield the same equations of
motion. Let $\hat{\mathcal{C}}$ denote the genus $(n-1)^2$ spectral
curve of the monopole and $\mathcal{C}$ denote the genus $n-1$
spectral curve of the Toda theory. We have already noted that
$\mathcal{C}=\hat{\mathcal{C}}/\texttt{C}_n$ and the natural
projection $\pi:\hat{\mathcal{C}}\rightarrow\mathcal{C}$ is an
$n$-fold unbranched cover. The solution of an integrable system is
typically expressed in terms of the straight line motion on the
Jacobian of the system's spectral curve. Such a line is determined
both by its direction and a point on the Jacobian. We shall now show
that both the direction and point relevant for monopole solutions
are obtained as pull-backs of Toda data.

First we recall that meromorphic differentials describe flows, and
that a meromorphic differential on a Riemann surface is uniquely
specified by its singular parts together with some normalisation
conditions. If
${\left\{\hat{\mathfrak{a}}_i,\hat{\mathfrak{b}}_i\right\}_{i=1}\sp{\hat
g}}$ form a  canonical basis for
$H_1(\hat{\mathcal{C}},\mathbb{Z})$,
    $$\hat{\mathfrak{a}}_i\cap\hat{\mathfrak{b}}_j
    =-\hat{\mathfrak{b}}_j\cap\hat{\mathfrak{a}}_i=\delta_{ij},$$
then one such normalisation condition is that the
$\hat{\mathfrak{a}}$-periods of the meromorphic differential vanish.
(Thus the freedom to add to the meromorphic differential a
holomorphic differential without changing its singular part is
eliminated.) In what follows we denote by
${\left\{\mathfrak{a}_i,\mathfrak{b}_i\right\}_{i=1}\sp{g}}$ a
similar canonical basis for $H_1(\mathcal{C},\mathbb{Z})$.

For the monopole the Lax operator
$A(\zeta)
$ has poles at $\zeta=\infty$. If we denote $\hat{\infty}_j$ to be
the $n$ points on the spectral curve above $\zeta=\infty$ (and these
may be assumed distinct) then we find that ${\eta}/{\zeta}=\rho_j
\zeta$ as $ \zeta\sim\hat{\infty}_j$. Consequently in terms of a
local coordinate
 $t$ at $\hat{\infty}_j$, $\zeta=1/{t}$, then
$$d\left(\frac{\eta}{\zeta}\right)=\left(-\frac{\rho_j}{t\sp2}+O(1)\right)dt.
$$
Thus on the monopole spectral curve we may uniquely define a
meromorphic differential by the pole behaviour at $\hat{\infty}_j$
and normalization
\begin{align*}
\gamma_\infty&=\left(\frac{\rho_j}{t\sp2}+O(1)\right)dt, \qquad
    0=\oint_{\hat{\mathfrak{a}_i}}\gamma_\infty.
\end{align*}
The vector of $\mathbf{\hat{\mathfrak{b}}}$-periods,
$$
    \boldsymbol{\widehat U} =\frac{1}{2i\pi}
\oint_{\mathbf{\hat{\mathfrak{b}}}} \gamma_\infty,$$ known as the
Ercolani-Sinha vector \cite{es89}, determines the direction of the
monopole flow on $\Jac(\hat{\mathcal{C}})$. This vector is in fact
constrained. Let us first recall Hitchin's conditions on a monopole
spectral curve, equivalent to the Nahm data already given. These are
\begin{itemize}\item[\bf{H1}] Reality conditions
$\displaystyle{
a_r(\zeta)=(-1)^r\zeta^{2r}\overline{a_r(-1/{\overline{\zeta}}\,)}}$

\item[\bf{H2}] Let  $L^{\lambda}$ denote the holomorphic line bundle on
    $T\PP\sp1$ defined by the transition function
    $g_{01}=\rm{exp}(-\lambda\eta/\zeta)$
and let
 $L^{\lambda}(m)\equiv
L^{\lambda}\otimes\pi\sp*\mathcal{O}(m)$ be similarly defined in
terms of the transition function
$g_{01}=\zeta^m\exp{(-\lambda\eta/\zeta)}$. Then $L^2$ is trivial on
$\hat{\mathcal{C}}$ and $L\sp1(n-1)$ is real.

\item[\bf{H3}] $H^0(\hat{\mathcal{C}},L^{\lambda}(n-2))=0$ for
    $\lambda\in(0,2)$
\end{itemize}
We have already seen the reality conditions. Here the triviality of
$L^2$ means that there exists a nowhere-vanishing holomorphic
section. The following are equivalent \cite{es89, hmr99}:
\begin{enumerate} \item $L\sp2$ is trivial on $\hat{\mathcal{C}}$.

\item $2\boldsymbol{\widehat U}\in \Lambda\Longleftrightarrow$ $
\boldsymbol{\widehat
U}=\frac{1}{2\pi\imath}\left(\oint_{\hat{\mathfrak{b}}_1}\gamma_{\infty},
\ldots,\oint_{\hat{\mathfrak{b}}_{\hat
g}}\gamma_{\infty}\right)\sp{T}= \frac12
\boldsymbol{n}+\frac12\hat{\tau}\boldsymbol{m} . $

\item There exists a 1-cycle
$\widehat{\mathfrak{es}}=\boldsymbol{n}\cdot{\hat{\mathfrak{a}}}+
\boldsymbol{m}\cdot{\hat{\mathfrak{b}}}$ such that for every
holomorphic differential
$$\Omega=\dfrac{\beta_0\eta^{n-2}+\beta_1(\zeta)\eta^{n
-3}+\ldots+\beta_{n-2}(\zeta)}{{\partial\mathcal{P}}/{\partial
\eta}}\,d\zeta, \qquad
\oint\limits_{\widehat{\mathfrak{es}}}\Omega=-2\beta_0. $$

\end{enumerate}
Here $\hat\tau$ is the period matrix of $\hat{\mathcal{C}}$ and
$\Lambda$ is the associated period lattice of the curve. Thus
$\boldsymbol{\widehat{U}}$ is constrained to be a half-period. These
are known as the Ercolani-Sinha constraints  and they impose $\hat
g$ \emph{transcendental constraints} on the curve yielding
$$\sum_{j=2}\sp{n}(2j+1)-\hat{g}=(n+3)(n-1)-(n-1)^2=4(n-1)$$
degrees of freedom.

We now turn to consider the behaviour of the Ercolani-Sinha vector
under a symmetry. Clearly our group acting the curve leads to an
action on divisors and consequently on the Jacobian. We now show
that the Ercolani-Sinha vector describing the flow is fixed under
the symmetry. This means the vector may be obtained from the
pull-back of a vector on the Jacobian of the quotient (Toda) curve.

Suppose we have a symmetry
$$0=P(\eta,\zeta)=P(\tilde\eta,\tilde\zeta)= \dfrac{{\tilde
P}(\eta,\zeta)}{(q\, \zeta+p)\sp{2n}}.$$ In particular
\begin{equation}
\partial_{\tilde\eta}P(\tilde\eta,\tilde\zeta)=(q\, \zeta+p)\sp{2}\partial_\eta
P(\tilde\eta,\tilde\zeta)=\frac{\partial_\eta {\tilde
P}(\eta,\zeta)}{(q\, \zeta+p)\sp{2n-2}}= \frac{\partial_\eta
{P}(\eta,\zeta)}{(q\, \zeta+p)\sp{2n-2}}.
\end{equation}
Using
$$d\tilde\zeta=\frac{d\zeta}{(q\, \zeta+p)\sp{2}}$$
we see then that
$$\frac{\tilde\zeta\sp{r}\tilde\eta\sp{s}d\tilde\zeta}{\partial_{\tilde\eta}P(\tilde\eta,\tilde\zeta)}=
\frac{(\bar p \zeta-\bar q)\sp{r}(q\,
\zeta+p)\sp{2n-4-r-2s}\eta\sp{s}d\zeta}{
\partial_\eta { P}(\eta,\zeta)}.$$
Bringing these together
\begin{lemma}The differential $\hat\omega_{r,s}=\dfrac{\zeta\sp{r}\eta\sp{s}d\zeta}{
\partial_\eta { P}(\eta,\zeta)}$ is invariant under the rotation
(\ref{pstransf}) if and only if
$$\zeta\sp{r}=(\bar p \zeta-\bar q)\sp{r}(q\,
\zeta+p)\sp{2n-4-r-2s}.$$ This always has a solution, the
holomorphic differential
$$\hat \omega=\dfrac{\eta\sp{n-2}d\zeta}{
\partial_\eta { P}(\eta,\zeta)}.$$
\end{lemma}
For the particular case of interest here, for rotations given by
$q=0$, $|p|\sp2=1$, then
\begin{equation}\label{phionhol}
\phi\sp\ast\left( \dfrac{\zeta\sp{r}\eta\sp{s}d\zeta}{
\partial_\eta { P}(\eta,\zeta)} \right)=\omega\sp{r+s+2}\,
\dfrac{\zeta\sp{r}\eta\sp{s}d\zeta}{
\partial_\eta { P}(\eta,\zeta)}
\end{equation}
and we also have solutions for each $s$ ($0\le s\le n-2$) and
$r=n-2-s$. These give us $g=n-1$ $\texttt{C}_n$-invariant
holomorphic differentials which are pullbacks of the holomorphic
differentials on $\mathcal{C}$. We remark also that the symmetry
always fixes the subspaces $\sum_r \mu_r \omega_{r,s}$ for fixed
$s$. Thus on the space of holomorphic differentials
$\{\hat\omega_I\}_{I=1}\sp{\hat g -1}\cup\{\hat\omega_{0,n-2}\}$
(for appropriate $I=(r,s)$ whose order does not matter) we have
\begin{equation}\label{symL}
\phi\sp\ast(\hat\omega_{1},\ldots,\hat\omega_{\hat g -1},\hat\omega_{0,n-2})=
(\hat\omega_{1},\ldots,\hat\omega_{\hat g -1},\hat\omega_{0,n-2})
\begin{pmatrix}\ast&\ast&0\\
\ast&\ast&0\\
0&0&1\end{pmatrix}:= (\hat\omega_{1},\ldots,\hat\omega_{\hat g -1},\hat\omega_{0,n-2})L
\end{equation}
where $L$ is a $\hat{g}\times \hat{g}$ complex matrix. As $L^n=1$,
the matrix is both invertible and diagonalizable.

With $\{\hat{\mathfrak{a}}_i,\hat{\mathfrak{b}}_i\}$ the canonical
homology basis introduced earlier and $\{\hat{u}_j\}$ a basis of
holomorphic differentials for our Riemann surface
$\hat{\mathcal{C}}$ we have the matrix of periods
\begin{equation}\begin{pmatrix}\oint_{\hat{\mathfrak{a}}_i}\hat{u}_j\\
\oint_{\hat{\mathfrak{b}}_i}\hat{u}_j\end{pmatrix}=
\begin{pmatrix}\hat{\mathcal{A}}\\
\hat{\mathcal{B}}\end{pmatrix}=\begin{pmatrix}1\\
\hat{\tau}\end{pmatrix}\hat{\mathcal{A}}
\end{equation}
with $\hat{\tau}=\hat{\mathcal{B}}\hat{\mathcal{A}}\sp{-1}$ the
period matrix. If $\sigma$ is any automorphism of
$\hat{\mathcal{C}}$ then $\sigma$ acts on
$H_1(\hat{\mathcal{C}},\mathbb{Z})$ and the holomorphic
differentials by
$$\sigma_\ast\begin{pmatrix}\hat{\mathfrak{a}}_i\\
\hat{\mathfrak{b}}_i\end{pmatrix}=\begin{pmatrix}{A}&B\\
C&D\end{pmatrix}\begin{pmatrix}\hat{\mathfrak{a}}_i\\
\hat{\mathfrak{b}}_i\end{pmatrix},\qquad
\sigma\sp\ast\hat{u}_j=\hat{u}_k
L\sp{k}_j,$$ where $\begin{pmatrix}{A}&B\\
C&D\end{pmatrix}\in Sp(2\hat{g},\mathbb{Z})$ and $L\in
GL(\hat{g},\mathbb{C})$. Then from
\begin{equation*}\oint_{\sigma_\ast\gamma}\hat{u}=\oint_\gamma\sigma\sp\ast\hat{u}
\end{equation*}
we obtain
\begin{align}
\begin{pmatrix}{A}&B\\
C&D\end{pmatrix}\begin{pmatrix}\hat{\mathcal{A}}\\
\hat{\mathcal{B}}\end{pmatrix}&=\begin{pmatrix}\hat{\mathcal{A}}\\
\hat{\mathcal{B}}\end{pmatrix}L .
\end{align}

With the ordering of holomorphic differentials of (\ref{symL}) the
second of the equivalent conditions for the Ercolani-Sinha vector
says there exist integral vectors $\mathbf{n}$, $\mathbf{m}$ such
that
\begin{equation}\label{escondab}
(\mathbf{n},\mathbf{m})\begin{pmatrix}\hat{\mathcal{A}}\\
\hat{\mathcal{B}}\end{pmatrix}=-2(0,\ldots,0,1).
\end{equation}
Now suppose $\sigma$ corresponds to a symmetry coming from a
rotation. Then the form of $L$ in (\ref{symL}) gives
\begin{align*}
(\mathbf{n},\mathbf{m})\begin{pmatrix}\hat{\mathcal{A}}\\
\hat{\mathcal{B}}\end{pmatrix}=-2(0,\ldots,0,1)=-2(0,\ldots,0,1).L=
(\mathbf{n},\mathbf{m})\begin{pmatrix}\hat{\mathcal{A}}\\
\hat{\mathcal{B}}\end{pmatrix}.L
=(\mathbf{n},\mathbf{m})\begin{pmatrix}{A}&B\\
C&D\end{pmatrix}\begin{pmatrix}\hat{\mathcal{A}}\\
\hat{\mathcal{B}}\end{pmatrix}
\end{align*}
and so
$$\left( (\mathbf{n},\mathbf{m})-(\mathbf{n},\mathbf{m})\begin{pmatrix}{A}&B\\
C&D\end{pmatrix}\right)\begin{pmatrix}\hat{\mathcal{A}}\\
\hat{\mathcal{B}}\end{pmatrix}=0.$$ As the rows of the lattice generated by $\begin{pmatrix}\hat{\mathcal{A}}\\
\hat{\mathcal{B}}\end{pmatrix}$ are independent over $\mathbb{Z}$ we
therefore have that
$$(\mathbf{n},\mathbf{m})=(\mathbf{n},\mathbf{m})\begin{pmatrix}{A}&B\\
C&D\end{pmatrix}$$ for all symplectic matrices $\begin{pmatrix}{A}&B\\
C&D\end{pmatrix}$ representing the symmetries coming from spatial
rotations. In particular $(\mathbf{n},\mathbf{m})$ is invariant
under the group of symmetries. Therefore the Ercolani-Sinha vector
is invariant and so as an element of the Jacobian, this will reduce
to a vector of the Jacobian of the quotient curve. Viewing this
vector as a divisor on the curve it projects to a divisor on the
quotient curve. Thus we have established

\begin{theorem} The Ercolani-Sinha vector is invariant under the
group of symmetries of the spectral curve arising from rotations
(\ref{pstransf}),
\begin{equation}
\boldsymbol{\widehat U}=\pi\sp\ast(\boldsymbol{ U}),\qquad
\boldsymbol{ U}\in\Jac(\mathcal{C}).
\end{equation}
\end{theorem}

For the cyclic symmetry under consideration we have from
\begin{align}dy&=n\left(\nu+\frac{(-1)\sp{n}|\beta|\sp2}{\nu}\right)\frac{d\zeta}{\zeta}=
-n(x\sp{n}+a_2 x\sp{n-2}+\ldots +a_n)\frac{d\zeta}{\zeta}, \nonumber \\
\partial_\eta { P}(\eta,\zeta)&=\zeta\sp{n-1}\partial_x
(x\sp{n}+a_2 x\sp{n-2}+\ldots +a_n), \nonumber \intertext{that}
\dfrac{\zeta\sp{n-2-s}\eta\sp{s}d\zeta}{
\partial_\eta { P}(\eta,\zeta)}&=
\pi\sp\ast\left(-\frac{1}{n}\,\frac{x\sp{s}dx}{y}\right).\label{pullbackhyp}
\end{align}
Thus each of the invariant differentials (for $0\le s\le n-2$)
reduce to hyperelliptic differentials.

\section{The base point}

In the construction of monopoles there is a distinguished point
$\widetilde{\boldsymbol{K}}\in\Jac(\hat{\mathcal{C}})$ that Hitchin
uses to identify degree $\hat{g}-1$ line bundles with
$\Jac(\hat{\mathcal{C}})$. For $n\ge3$ this point is a singular
point of the theta divisor, $ \widetilde{\boldsymbol{K}}\in
\Theta_{\rm singular}$ \cite{bren06}. If we denote the Abel map by
$$\mathcal{A}_{\hat{Q}}(\hat{P})=\int_{\hat{Q}}\sp{\hat{P}}\hat{u}_i$$
then \begin{equation}\label{tildeK}\widetilde{\boldsymbol{K}}=
\boldsymbol{\hat{K}}_{\hat{Q}}+\mathcal{A}_{\hat{Q}}\left((n-2)
\sum_{k=1}\sp{n}\hat{\infty}_k\right).\end{equation} Here
$\boldsymbol{\hat{K}}_{\hat{Q}}$ is the vector of Riemann
constants for the curve $\hat{\mathcal{C}}$. If
$\mathcal{K}_{\hat{\mathcal{C}}}$  is the canonical divisor of the
curve then
$\mathcal{A}_{\hat{Q}}(\mathcal{K}_{\hat{\mathcal{C}}})=-2\boldsymbol{\hat{K}}_{\hat{Q}}$.
The righthand side of (\ref{tildeK}) is in fact independent of the
base point $\hat{Q}$ in its definition.

The point $\widetilde{\boldsymbol{K}}$ is the base point of the
linear motion in the Jacobian referred to earlier and we shall now
relate this to a point in the Jacobian of the Toda spectral curve
$\mathcal{C}$. Let $\mathcal{A}_Q(\mathcal{K}_\mathcal{C})=-2{\boldsymbol{K}}_Q$ be
the corresponding quantities for the curve $\mathcal{C}$ with basis
of holomorphic differentials $\{ u_a\}$. We first relate $\pi\sp\ast
{\boldsymbol{K}}_Q$ and ${\hat {\boldsymbol{K}}}_{\hat Q}$ where $\pi({\hat Q})=Q$ is some
preimage of $Q$. Let our symmetry be
$\phi:\hat{\mathcal{C}}\rightarrow \hat{\mathcal{C}}$,
$\phi\sp{n}=1$, and observe that (with $\pi({\hat P})=P$, $\pi({\hat
Q})=Q$)
\begin{align*}\pi\sp\ast(\mathcal{A}_Q(P))&=\pi\sp\ast\left(\int_Q\sp{P}u \right)
=\sum_{s=0}\sp{n-1}\int_{\phi\sp{s}(\hat Q)}\sp{\phi\sp{s}(\hat
P)}\hat u =\sum_{s=0}\sp{n-1}\left[\mathcal{A}_{\hat
Q}\left(\phi\sp{s}(\hat P)\right)-\mathcal{A}_{\hat
Q}\left(\phi\sp{s}(\hat Q)\right)\right].
\end{align*}
(This is actually independent of the base-point chosen for the Abel
map, so well-defined.) Now if $\sum_{\alpha=1}\sp{2g-2}P_\alpha$ is
a canonical divisor for $\mathcal{C}$ then
$\sum_{\alpha=1}\sp{2g-2}\sum_{s=0}\sp{n-1}\phi\sp{s}(\hat
P_\alpha)$ is a canonical divisor for $\hat{\mathcal{C}}$. Thus
\begin{align*}
\pi\sp\ast(-2{\boldsymbol{K}}_Q)&=\pi\sp\ast\left(\mathcal{A}_Q(\mathcal{K}_\mathcal{C})
\right)\\
&=\pi\sp\ast\left(\sum_{\alpha=1}\sp{2g-2}\int_Q\sp{P_\alpha}u
\right)\\
&=\mathcal{A}_{\hat Q}(\hat{\mathcal{K}}_\mathcal{\hat{C}})-2(g-1)
\sum_{s=0}\sp{n-1}\mathcal{A}_{\hat Q}\left(\phi\sp{s}(\hat
Q)\right)\\
&=-2{\hat {\boldsymbol{K}}}_{\hat Q}-2(g-1) \sum_{s=0}\sp{n-1}\mathcal{A}_{\hat
Q}\left(\phi\sp{s}(\hat Q)\right).
\end{align*}
Therefore
\begin{equation}\label{pullbackK}
\pi\sp\ast({\boldsymbol{K}}_Q)={\hat {\boldsymbol{K}}}_{\hat Q}+(g-1)
\sum_{s=0}\sp{n-1}\mathcal{A}_{\hat Q}\left(\phi\sp{s}(\hat
Q)\right)+\hat e,
\end{equation}
where $2\hat e\in \Lambda$ is a half-period. This expression may be
rewritten as
\begin{align*}
\pi\sp\ast({\boldsymbol{K}}_Q)&={\hat {\boldsymbol{K}}}_{\hat Q}+(g-1)
\sum_{s=0}\sp{n-1}\mathcal{A}_{\hat Q}\left(\phi\sp{s}(\hat
Q)\right)+\hat e\\
&=\left[{\hat {\boldsymbol{K}}}_{\hat Q}+(\hat g-1)\mathcal{A}_{\hat Q}(\hat
P)\right]-(\hat g-1)\mathcal{A}_{\hat Q}(\hat P)+(g-1)
\sum_{s=0}\sp{n-1}\mathcal{A}_{\hat Q}\left(\phi\sp{s}(\hat
Q)\right)+\hat e\\
&={\hat {\boldsymbol{K}}}_{\hat P}-n( g-1)\mathcal{A}_{\hat Q}(\hat P)+(g-1)
\sum_{s=0}\sp{n-1}\mathcal{A}_{\hat Q}\left(\phi\sp{s}(\hat
Q)\right)+\hat e\\
&={\hat {\boldsymbol{K}}}_{\hat P}+(g-1)\sum_{s=0}\sp{n-1}\mathcal{A}_{\hat
P}\left(\phi\sp{s}(\hat Q)\right)+\hat e,
\end{align*}
showing the left-hand side is independent of the choice of
base-point for the Abel map.

Comparison of (\ref{tildeK}) and (\ref{pullbackK}) now shows that
\begin{equation}\label{tildeKpb}\widetilde{\boldsymbol{K}}=
\pi\sp\ast({\boldsymbol{K}}_{\infty_+})-\hat e\end{equation} where
$\pi(\hat{\infty}_k)=\infty_+$ as noted earlier. Now the half-period
$\hat e$ can be identified and is of the form $\hat
e=\pi\sp\ast(e)$. The actual identification depends on an homology
choice and will be given in the next section, but for the moment we
simply note the form
\begin{equation}\label{tildeKpbf}\widetilde{\boldsymbol{K}}=
\pi\sp\ast({\boldsymbol{K}}_{\infty_+}-e).\end{equation}

\section{Fay-Accola factorization}

The standard reconstruction of solutions for an integrable system with spectral
curve $\hat{\mathcal{C}}$ proceeds by constructing the Baker-Akhiezer functions
for this curve.
These may be calculated in terms of theta functions for the curve and for our
present purposes we may focus on the theta function
$\theta(\lambda\widehat{\boldsymbol{U}}-
\widetilde{\boldsymbol{K}}\,|\,\hat\tau)$. This describes a flow on the Jacobian of
$\hat{\mathcal{C}}$ in the direction of the Ercolani-Sinha vector $\widehat{\boldsymbol{U}}$
with base point $\widetilde{\boldsymbol{K}}$. We have observed that we have
a cyclic unramified covering $\pi:\hat{\mathcal{C}}\rightarrow\mathcal{C}$
of the affine Toda spectral curve by the monopole spectral curve.
The map $\pi$ leads to a map $\pi\sp\ast:
\text{Jac}({\mathcal{C}})\rightarrow\text{Jac}(\hat{\mathcal{C}})$
which may be lifted to $\pi\sp\ast:{\mathbb C}\sp{g}\rightarrow
{\mathbb C}\sp{\hat g}$. Further we have established that
$$\lambda\widehat{\boldsymbol{U}}-
\widetilde{\boldsymbol{K}}=
\pi\sp\ast(\lambda \boldsymbol{U} -{\boldsymbol{K}}_{\infty_+}+e).
$$
We now are in a position to make use of a remarkable factorization theorem due to
Accola and Fay \cite{acc71, fay73} and also observed by Mumford.
When $\hat z=\pi\sp\ast z$ the theta
functions on $\hat{\mathcal{C}}$ and ${\mathcal{C}}$ are related
by this factorization theorem,
\begin{theorem}[\bf Fay-Accola] \label{fayaccola}
With respect to the ordered canonical homology bases $\{
\hat{\mathfrak{a}}_i\sp{c},\hat{\mathfrak{b}}_i\sp{c}\}$ described
below and for arbitrary $\boldsymbol{ z}=\in \mathbb{C}^g$ we have
\begin{equation}
\frac{\theta[\hat e](\pi\sp\ast\boldsymbol{ z};\hat{\tau}\sp{c})}
{\prod_{k=0}^{n-1}\theta\left[\begin{matrix}0&0&\dots&0
\\ \frac{k}{n}&0&\dots&0 \end{matrix}\right]\left(\boldsymbol{ z};\tau\sp{c}
\right)}
=c_0(\widehat{\tau}\sp{c}
) \label{fafactora}
\end{equation}
is a non-zero modular constant
$c_0(\hat{\tau}\sp{c}
)
$ independent of
$\boldsymbol{ z}$. Here $\hat{\tau}\sp{c}$ is the $\mathfrak{a}$-normalized
period matrix for the curve $\hat{\mathcal{C}}$ in this homology basis and
$$\hat e=\left[\begin{matrix}0&0&\dots&0
\\ \frac{n-1}{2}&0&\dots&0 \end{matrix}\right]=
\pi\sp\ast\left(e\right)=
\pi\sp\ast\left( \left[\begin{matrix}0&0&\dots&0
\\ \frac{n-1}{2n}&0&\dots&0 \end{matrix}\right]\right)
.$$
\end{theorem}
The significance of this theorem for our setting is that it means
we can reduce the construction of solutions to that of quantities
purely in terms of the hyperelliptic affine Toda spectral curve.

The theorem is expressed in terms of a particular choice of homology
basis which is well adapted to the symmetry at hand. In terms of the
conformal automorphism
$\phi:\hat{\mathcal{C}}\rightarrow\hat{\mathcal{C}}$ of
$\hat{\mathcal{C}}$ that generates the group
$\texttt{C}_n=\{\phi\sp{s}\,|\,0\le s\le n-1\}$ of cover
transformations of $\hat{\mathcal{C}}$ and the projection
$\pi:\hat{\mathcal{C}}\rightarrow\mathcal{C}$ there exists a basis
 $\{\hat{\mathfrak{a}}_0\sp{c},\hat {\mathfrak{b}}_0\sp{c},
 \hat{\mathfrak{a}}_1\sp{c},\hat
 {\mathfrak{b}}_1\sp{c},\ldots,\hat
{\mathfrak{a}}_{\hat g-1}\sp{c},\hat {\mathfrak{b}}_{\hat
g-1}\sp{c}\}$ of homology cycles for $\hat{\mathcal{C}}$ and
$\{{\mathfrak{a}}_0\sp{c},{\mathfrak{b}}_0\sp{c},{\mathfrak{a}}_1\sp{c},
{\mathfrak{b}}_1\sp{c},\ldots,{\mathfrak{a}}_{g-1}\sp{c},{\mathfrak{b}}_{g-1}\sp{c}\}$
for $\mathcal{C}$ such that ( for $1\le j\le g-1,\ 0\le s\le n$)
\begin{align*}
\pi(\hat
{\mathfrak{a}}_0\sp{c})&={\mathfrak{a}}_0\sp{c},&\pi(\hat
{\mathfrak{a}}_{j+s(g-1)}\sp{c})&={\mathfrak{a}}_j\sp{c},&
\pi(\hat {\mathfrak{b}}_0\sp{c})&=n\,
{\mathfrak{b}}_0\sp{c},&\pi(\hat
{\mathfrak{b}}_{j+s(g-1)}\sp{c})&={\mathfrak{b}}_j\sp{c},&\\
\phi\sp{s}(\hat {\mathfrak{a}}_0\sp{c})&\sim {\hat
{\mathfrak{a}}_0\sp{c}},&\phi\sp{s}(\hat
{\mathfrak{a}}_j\sp{c})&=\hat
{\mathfrak{a}}_{j+s(g-1)}\sp{c},&
\phi\sp{s}(\hat  {\mathfrak{b}}_0)&= {\hat
{\mathfrak{b}}_0\sp{c}},&\phi\sp{s}(\hat
{\mathfrak{b}}_j\sp{c})&=\hat {\mathfrak{b}}_{j+s(g-1)}\sp{c}.&
\end{align*}
Here $\phi\sp{s}(\hat {\mathfrak{a}}_0)$ is homologous to $\hat
{\mathfrak{a}}_0$. If $\hat v_i$ are the
$\hat{\mathfrak{a}}$-normalized differentials for
$\hat{\mathcal{C}}$, then
$$\delta_{i,j+s(g-1)}=\int_{\hat {\mathfrak{a}}_{j+s(g-1)}}\hat v_i=
\int_{\phi\sp{s}(\hat {\mathfrak{a}}_j)}\hat v_i=\int_{\hat
{\mathfrak{a}}_j}(\phi\sp{s})\sp{\ast}\hat v_i=\int_{\hat
{\mathfrak{a}}_j}\hat v_{i-s(g-1)},$$ and we find that
\begin{equation}(\phi\sp{s})\sp{\ast}\hat v_0= \hat
v_0,\qquad(\phi\sp{s})\sp{\ast}\hat v_i=\hat v_{i-s(g-1)}.
\label{gpdiff} \end{equation} If $v_i$ are the normalized
differentials for $\mathcal{C}$, then
$$
\delta_{ij}=\int_{{\mathfrak{a}}_j}v_i=\int_{\pi(\hat
{\mathfrak{a}}_{j+s(g-1)})}v_i
=\int_{{\mathfrak{a}}_{j+s(g-1)}}\pi\sp\ast (v_i)$$ shows that
$$\pi\sp\ast (v_i)=\hat v_i+(\phi)\sp{\ast}\hat v_i+\ldots
+(\phi\sp{p-1})\sp{\ast}\hat v_i$$ and similarly that
$$\pi\sp\ast (v_0)=\hat v_0.$$
We may use the characters of $\texttt{C}_n$ to construct the
remaining linearly independent differentials on
$\hat{\mathcal{C}}$.

From (\ref{gpdiff}) we have an action of $\texttt{C}_n$ on
$\text{Jac}(\hat{\mathcal{C}})$ which lifts to an automorphism of
${\mathbb C}\sp{\hat g}$ by
\begin{equation}\label{symmjac}
\phi\sp{s}(\hat z)=(\hat z_0,\hat z_{1-s(g-1)},\ldots,\hat
z_{g-1-s(g-1)},\ldots,\hat z_{1+(p-s-1)(g-1)},\ldots,\hat
z_{g-1+(p-s-1)(g-1)})
\end{equation}
Now (\ref{symmjac}) together with the invariance of the
Ercolani-Sinha vector mans that in this cyclic homology basis we
have
\begin{equation}\label{escyclic}
(\mathbf{n},\mathbf{m})=(r_0,\mathbf{r},\ldots,\mathbf{r},
s_0,\mathbf{s},\ldots,\mathbf{s})\end{equation}
 where the vectors
$\mathbf{r}= (r_1,\ldots,r_{g-1})$ and similarly $\mathbf{s}$ are
each repeated $n$ times. We also have
\begin{align}
\pi_\ast(\widehat{\mathfrak{es}} )&
= r_0\mathfrak{a}_0+n
\mathbf{r}\cdot\boldsymbol{ \mathfrak{a}}+ns_0\mathfrak{b}_0+ n
\mathbf{s}\cdot\boldsymbol{ \mathfrak{b}} . \label{projhomes}
\end{align}

With the choices above (things are different for
$\hat{\mathfrak{b}}$-normalization) we may lift the map
$\pi\sp\ast: \Jac(\mathcal{C})\rightarrow \Jac(\hat{\mathcal{C}})$
to $\pi\sp\ast:{\mathbb C}\sp{g}\rightarrow {\mathbb C}\sp{\hat
g}$,
$$
\pi\sp\ast(z)= \pi\sp\ast(z_0,z_1,\ldots,z_{g-1})=
(n\,z_0,z_1,\ldots,z_{g-1},\ldots,z_1,\ldots,z_{g-1})=\hat z.$$ With
this homology basis the period matrices for the two curves are
related by the block form
$$
\hat\tau\sp{c}=\left(
\begin{array}{ccccc}
n\, \tau_{00}\sp{c}&\tau_{0j}\sp{c}&\tau_{0j}\sp{c}&\ldots&\tau_{0j}\sp{c}\\
\tau_{j0}\sp{c}&\mathcal{M}&\mathcal{M}\sp{(1)}&&\mathcal{M}\sp{(n-1)}\\
\vdots&\\
\tau_{j0}\sp{c}&\mathcal{M}\sp{(1)}&&&\mathcal{M}
\end{array}
\right)
$$
where $\mathcal{M}\sp{(s)}=\int_{\phi\sp{-s}(\hat
{\mathfrak{b}}_j)}\hat v_i$. The $(r,s)$ block here has entry
$\mathcal{M}\sp{s-r}$ and
$(\mathcal{M}\sp{(s-r)})\sp{T}=\mathcal{M}\sp{(r-s)}$ by the
bilinear identity. Then
$\tau\sp{c}_{ij}=\sum_{s=0}\sp{n-1}\mathcal{M}\sp{(s)}_{ij}$. The
case $n=3$ is instructive, for here the $n-2$ block matrices are
just numbers and we have
\begin{equation}\hat{\tau}\sp{c}
=\left( \begin{array}{cccc} a&b&b&b\\
b&c&d&d\\
b&d&c&d\\
b&d&d&c
\end{array}\right),\qquad
\tau\sp{c}
=\left( \begin{array}{cc} \frac13 a&b\\
b&c+2d
\end{array}\right) .\end{equation}
The point to note is that although the period matrix for
$\hat{\mathcal{C}}$ involves integrations of differentials that do
not reduce to hyperelliptic integrals, the combination of terms
appearing in the reduction can be expressed in terms of
hyperelliptic integrals. This is a definite simplification.
Further the $\Theta$ function defined by $\hat\tau\sp{c}$ has the
symmetries
$$\Theta(\hat z|\hat\tau\sp{c})=\Theta(\phi\sp{s}(\hat z)|\hat\tau\sp{c})$$
for all $\hat z\in{\mathbb C}\sp{\hat g}$. In particular, the
$\Theta$ divisor is fixed under $\texttt{C}_n$.

If we are to reduce the construction of cyclic monopoles to a
problem involving only hyperelliptic quantities we must describe
the Ercolani-Sinha constaints in the context of the curve
$\mathcal{C}$.

\begin{theorem} \label{ESHYP} The Ercolani-Sinha constraint on the curve
$\hat{\mathcal{C}}$ yields the constraint
\begin{equation}
-2(0,\ldots,0,1)=(r_0,n\mathbf{r},n s_0,
n\mathbf{s})\begin{pmatrix}\mathcal{A}\\
\mathcal{B}\end{pmatrix}
\end{equation}
on the curve $\mathcal{C}$ with respect to the differentials
$u_s=-{x\sp{s}dx}/{(ny)}$ ($s=0,\ldots,n-2$).
\end{theorem}
\begin{proof}
The invariance of the Ercolani-Sinha vector means that
$\phi\sp\ast(\widehat{\mathfrak{es}})=\widehat{\mathfrak{es}}$.
Thus
$$
\int_{\widehat{\mathfrak{es}} }\hat\omega_{r,s}=
\int_{\phi\sp\ast(\widehat{\mathfrak{es}})}\hat\omega_{r,s}=
\int_{\widehat{\mathfrak{es}} }\phi\sp\ast\hat\omega_{r,s}=
\omega\sp{r+s+2}\int_{\widehat{\mathfrak{es}} }\hat\omega_{r,s},
$$
where we have used (\ref{phionhol}). Thus the integral of any
noninvariant differential around the cycle
$\widehat{\mathfrak{es}}$ must vanish, while from
(\ref{pullbackhyp}) and the Ercolani-Sinha condition we have that
$$-2\,\delta_{s,n-2}=\int_{\widehat{\mathfrak{es}} }
\pi\sp\ast\left(-\frac{1}{n}\,\frac{x\sp{s}dx}{y}\right)=
\int_{\pi_\ast(\widehat{\mathfrak{es}} )}
-\frac{1}{n}\,\frac{x\sp{s}dx}{y}.
$$
The theorem then follows upon using (\ref{projhomes}).
\end{proof}

In actual calculations it is convenient to use the unnormalized
differentials $\hat\omega_{r,s}$ and ${x\sp{s}dx}/{(ny)}$ rather
than Fay's normalized differentials $\hat v_i$. An alternate proof
of Theorem \ref{ESHYP} via Poincar\'e's reducibility theorem is
given in the Appendix, which provides further useful relations
amongst the periods of the two curves.

\section{Discussion}
In this paper we have shown that any cyclically symmetric monopole
is gauge equivalent to Nahm data obtained via Sutcliffe's ansatz
from the affine Toda equations. Further, the data needed to
reconstruct the monopole, the Ercolani-Sinha vector and base point
for linear flow on the Jacobian, may also be obtained from data on
the affine Toda equation's hyperelliptic spectral curve
$\mathcal{C}$. A theorem of Fay and Accola then enables us to
express the theta functions for the monopole spectral curve in
terms of the theta functions for the curve $\mathcal{C}$. Finally
the transcendental constraints on the monopole's spectral curve
can be recast as transcendental constraints for the hyperelliptic
curve $\mathcal{C}$ (Theorem \ref{ESHYP}). At this stage then the
construction of cyclically symmetric monopoles has been reduced to
one entirely in terms of hyperelliptic curves. Although analogues
of both the transcendental constraints still exist this is a
significant simplification. We note that the structure of the
theta divisor is better understood in the hyperelliptic setting
\cite{vanh95} and the hyperelliptic integrals are somewhat simpler
than the general integrals appearing in the Ercolani-Sinha
constraint for the full monopole curve.

Other approaches to constructing monopoles are known. In
particular \cite{hmm95} describe cyclically symmetric monopoles
within the rational map approach (see also \cite[\S8.8]{ms04}).
These authors show that the rational map for monopoles with
$\texttt{C}_n$ invariance about the $x_3$-axis takes the form
$$R(z)=\frac{\mu z\sp{l}}{z\sp{n}-\nu}$$
where $0\le l\le n-1$. The complex quantity $\nu $ determines
$\mu$ when the monopoles are strongly centred. Here
$\nu=(-1)\sp{n-1}\bar\beta$ of equation (\ref{cycliccurve}). The
moduli space $\mathcal{M}_n\sp{l}$ is a 4-dimensional totally
geodesic submanifold of the full moduli space. It is interesting
that both the rational map description and the description we have
presented lead to extra discrete parameters ($l$ in the case of
rational maps, and $k$ in \ref{fafactora}). The connection, if
any, between these will be pursued elsewhere \cite{bde10}.

Clearly the ansatz for monopoles extends to other algebras. If we
construct the spectral curve from the $D_n$ Toda system using the
$2n$ dimensional representation we find a spectral curve
$\hat{\mathcal{C}}$ of the form
$$\eta^{2n} + a_1 \eta^{2n-2} \zeta^2 + a_2\eta^{2n-4} \zeta^4+
    \ldots+ a_n\zeta^{2n}+\alpha\eta^2 (\frac1{w} + \zeta^{4n-4}w)
    =0.$$
Letting $x=\eta/\zeta$ the curve (upon dividing by $\zeta\sp{2n}$)
becomes
$$x^{2n} + a_1 x^{2n-2}  + a_2 x^{2n-4}+
    \ldots+ a_n+\alpha x^2(\frac1{w\zeta^{2n-2}} + \zeta^{2n-2}w)
    =0.$$
and so we get with $\nu=\alpha w\zeta^{2n-2}$
$$P_n(x^2)+x^2(\nu+\frac{\alpha\sp2}{\nu})
    =0$$
leading to a hyperelliptic curve $\tilde{\mathcal{C}}$
$$y^2=P_n(x^2)^2-4\alpha\sp2 x^4.$$
This curve has cyclic symmetry $\texttt{C}_{2n-2}$ from the
appearance of $\zeta^{2n-2}$ and $\texttt{C}_2$ due to the
appearance of $x^2$. The genus of $\hat{\mathcal{C}}$ is
$(2n-1)^2-2n$. The genus of $\tilde{\mathcal{C}}$ is $2n-1$.
Finally $\tilde{\mathcal{C}}$ covers a genus $n-1$ curve
${\mathcal{C}}$
$$y^2=P_n(u)^2-4\alpha\sp2 u^2.$$
Here we expect the Toda motion to lie in the Prym of this
covering, but the general theory warrants further study.

\section*{Acknowledgements}
I have benefited from many discussions with Antonella D'Avanzo,
Victor Enolskii and Timothy P. Northover. The results presented
here were described at the MISGAM supported meeting ``From
Integrable Structures to Topological Strings and Back'', Trieste
2008, and the Lorentz Center meeting ``Integrable Systems in
Quantum Theory'', Leiden 2008. I am grateful to the organisers of
these meetings for providing such a stimulating and pleasant
environment.

\appendix

\section{Proof of Theorem \ref{ESHYP} via Poincar\'e Reducibility}

It is instructive to see an alternative proof of Theorem
\ref{ESHYP} in terms of Poincar\'e's reducibility condition, which
we now recall. Consider Riemann matrices
$$\hat\Pi=
\begin{pmatrix}\hat{\mathcal{A}}\\
\hat{\mathcal{B}}\end{pmatrix}=\begin{pmatrix}1\\
\hat{\tau}\end{pmatrix}\hat{\mathcal{A}},\qquad
\Pi=
\begin{pmatrix}\mathcal{A}\\
\mathcal{B}\end{pmatrix}=\begin{pmatrix}1\\
\tau\end{pmatrix}\mathcal{A},
$$
where $\hat{\mathcal{A}}$ and $\hat{\mathcal{B}}$ are the $\hat
g\times \hat g$ matrices of $\hat{\mathfrak{a}}$-periods and
$\hat{\mathfrak{b}}$-periods respectively for the curve
$\hat{\mathcal{C}}$ with similarly named quantities for the curve
$\mathcal{C}$. If $\{\hat\gamma_a\}_{a=1}\sp{\hat 2g}$ is a basis
for $H_1(\hat{\mathcal{C}},\mathbb{Z})$,
$\{\hat\omega_\mu\}_{\mu=1}\sp{\hat g}$ a basis of holomorphic
differentials of $\hat{\mathcal{C}}$, and $\{\gamma_i\}_{i=1}\sp{
2g}$ a basis for $H_1(\mathcal{C},\mathbb{Z})$,
$\{\omega_\alpha\}_{\alpha=1}\sp{ g}$ a basis of holomorphic
differentials of ${\mathcal{C}}$, these are related by
\begin{align*}
\pi_\ast(\hat\gamma_a)=M_a\sp{\ i}\,\gamma_i,\qquad
\pi\sp{\ast}(\omega_\mu)=\hat\omega_\alpha \,\lambda\sp{\alpha}_{\ \mu}.
\end{align*}
Here $\lambda$ is complex $ {\hat g}\times g$-matrix of maximal
rank and $M$ is a $ 2{\hat g}\times 2 g$-matrix of integers of
maximal rank. Then from
\begin{equation*}(M\Pi)_{a\mu}=M_a\sp{\ i}\oint_{\gamma_i}\omega_\mu=
\oint_{\pi_\ast{\hat\gamma}_a}\omega_\mu=
\oint_{{\hat\gamma}_a}\pi\sp\ast\omega_\mu=\oint_{{\hat\gamma}_a}
{\hat\omega}_\alpha \,\lambda\sp{\alpha}_{\ \mu}=(\hat\Pi\lambda)_{a\mu}
\end{equation*}
we obtain Poincar\'e's reducibility condition
\begin{equation}
 \hat{\Pi}{\lambda}=M\Pi . \label{reducibility}
\end{equation}
For the cyclic homology basis and corresponding
$\hat{\mathfrak{a}}$-normalized differentials $\hat v_i$ of Fay
this takes the form
$$
\begin{pmatrix}1\\
\hat{\tau}\sp{c}\end{pmatrix}{\mathcal I}' =
\begin{pmatrix}{\mathcal I}'&0\\0&{\mathcal I}\end{pmatrix}
\begin{pmatrix}1\\
\tau\sp{c}\end{pmatrix}:=M\begin{pmatrix}1\\
\tau\sp{c}\end{pmatrix},$$ where we define the ${\hat g}\times g$
matrices ${\mathcal I}$, ${\mathcal I}'$ and (to be used shortly)
$P$,
$${\mathcal I}=\left(\begin{matrix}n&0\\ 0&1_{g-1}\\ \vdots&\vdots\\
0&1_{g-1}\end{matrix}\right),\quad
{\mathcal I}'=\left(\begin{matrix}1&0\\ 0&1_{ g-1}\\ \vdots&\vdots\\
0&1_{g-1}\end{matrix}\right),\quad
P=\left(\begin{matrix}1_{g}\\ 0&\\ \vdots \\ 0\end{matrix}\right).
$$
For the same cyclic homology basis but an arbitrary basis of
holomorphic differentials we obtain (\ref{reducibility}) with
$$\lambda={\hat{\mathcal{A}}}\sp{-1} {\mathcal I}' \mathcal{A}.$$

Now bringing together the Ercolani-Sinha constraint
(\ref{escondab}) with (\ref{reducibility}) we find
\begin{equation*}
-2(0,\ldots,0,1)\lambda=
(\mathbf{n},\mathbf{m})\begin{pmatrix}\hat{\mathcal{A}}\\
\hat{\mathcal{B}}\end{pmatrix}\lambda=
(\mathbf{n},\mathbf{m})M\begin{pmatrix}\mathcal{A}\\
\mathcal{B}\end{pmatrix}=
(r_0,n\mathbf{r},n s_0, n\mathbf{s})\begin{pmatrix}\mathcal{A}\\
\mathcal{B}\end{pmatrix}.
\end{equation*}
where we have used (\ref{escyclic}) and that
$(\mathbf{n},\mathbf{m})M=(r_0,n\mathbf{r},n s_0, n\mathbf{s})$.
Here $\hat{\mathcal{A}}$ has been constructed from the
differentials $\hat\omega_{r,s}={\zeta\sp{r}\eta\sp{s}d\zeta}/{
\partial_\eta { P}(\eta,\zeta)}$ (which are not
Fay's normalized differentials $\hat v_i$) while the differentials
for $\mathcal{A}$ are as yet unspecified and we wish to construct
$\lambda$. Using (\ref{pullbackhyp}) it is convenient to choose
$u_s=-{x\sp{s}dx}/{(ny)}$ (so that $
\pi\sp\ast\left(u_s\right)=\hat\omega_{n-2-s,s} $) and to order
the differentials with the noninvariant differentials before the
invariant differentials, $\{\hat\omega_{r,s}\}_{r+s\ne
n-2}\cup\{\hat{\omega}_{n-2,0},\ldots,\hat{\omega}_{0,n-2} \}$.
Then we find the matrix of periods
$$\hat{\mathcal{A}}
=
\begin{pmatrix}
0&\ldots&0&\ast&\ldots&\ast\\
&\mathcal{D}\sp{(0)}&&&\mathcal{A}'&\\
&\mathcal{D}\sp{(1)}&&&\mathcal{A}'&\\
&\vdots&&&\vdots&\\
&\mathcal{D}\sp{(n-1)}&&&\mathcal{A}'&
\end{pmatrix}.
$$
Here the first row has zero entries for the periods of the
noninvariant differentials over the invariant cycle
$\hat{\mathfrak{a}}_0$ while $\mathcal{D}\sp{(k)}$ is the
$(g-1)\times(\hat g-g)$ matrix of periods of the noninvariant
differentials over the cycles $\hat{\mathfrak{a}}_{i+k(g-1)}$ (
$i=1,\ldots,g-1$). Thus
$$\mathcal{D}\sp{(k)}_{i,(r,s)}=
\int_{\mathfrak{a}_{i+k(g-1)}}\hat\omega_{r,s}=
\int_{\phi\sp{k}(\mathfrak{a}_{i})}\hat\omega_{r,s}=
\int_{\mathfrak{a}_{i}}(\phi\sp{k})\sp\ast\hat\omega_{r,s}=
\omega\sp{k(r+s+2)}\int_{\mathfrak{a}_{i}}\hat\omega_{r,s}=
\omega\sp{k(r+s+2)}\mathcal{D}\sp{(0)}_{i,(r,s)}.
$$
The matrix of periods $\mathcal{A}'$ of the invariant
differentials over the same cycles is such that
$$\int_{\hat{\mathfrak{a}_i}}\hat{\omega}_{n-2-s,s}=
\int_{\hat{\mathfrak{a}_i}}\pi\sp{\ast}( u_s)=
\int_{\pi_{\ast}(\hat{\mathfrak{a}_i})} u_s= \int_{{\mathfrak{a}_i}}
u_s,
$$
and the matrix of periods $\mathcal{A}$ for the curve
$\mathcal{C}$ appearing above is precisely the submatrix
$${\mathcal{A}}
=
\begin{pmatrix}
\ast&\ldots&\ast\\
&\mathcal{A}'&
\end{pmatrix}.
$$

Next we note that we may write \begin{align*}(0,\ldots,0,1)\lambda
\mathcal{A}\sp{-1}&= (0,\ldots,0,1)\hat{\mathcal{A}}\sp{-1}
{\mathcal I}'= (0,\ldots,0,1)\hat{\mathcal{A}}\sp{-1} C P
=(0,\ldots,0,1)\left( C\sp{-1}\hat{\mathcal{A}}\right)\sp{-1} P\\
&=\left( \left( C\sp{-1}\hat{\mathcal{A}}\right)\sp{-1}_{\hat
g,1},\ldots,\left( C\sp{-1}\hat{\mathcal{A}}\right)\sp{-1}_{\hat
g,g}\right)
\end{align*}
with
$$C=\left(\begin{matrix}1&0&0&\ldots&0\\
 0&1_{ g-1}&0&\ldots
&0\\
0&1_{ g-1}&1_{ g-1}&\ldots
&0\\ \vdots&\vdots& &\ddots&\\
0&1_{g-1}&0&\ldots&1_{ g-1}\end{matrix}\right), \qquad
C\sp{-1}=\left(\begin{matrix}1&0&0&\ldots&0\\
 0&1_{ g-1}&0&\ldots
&0\\
0&-1_{ g-1}&1_{ g-1}&\ldots
&0\\ \vdots&\vdots& &\ddots&\\
0&-1_{g-1}&0&\ldots&1_{ g-1}\end{matrix}\right).
$$
This factorization was motivated by the observation that
$$C\sp{-1}\hat{\mathcal{A}}
=
\begin{pmatrix}
0&\ldots&0&\ast&\ldots&\ast\\
&\mathcal{D}\sp{(0)}&&&\mathcal{A}'&\\
& \mathcal{D}\sp{(1)}-\mathcal{D}\sp{(0)}&&&0&\\
&\vdots&&&\vdots&\\
&\mathcal{D}\sp{(n-1)}-\mathcal{D}\sp{(n-2)}&&&0&
\end{pmatrix}=
\begin{pmatrix}\mathcal{E}&\mathcal{A}\\
\mathcal{F}&0
\end{pmatrix}
$$
and so upon noting that $\hat g-g$ is even and
$|C\sp{-1}\hat{\mathcal{A}}|=|\mathcal{A}|\,|\mathcal{F}|$ we have
the cofactor expression
$$\left( C\sp{-1}\hat{\mathcal{A}}\right)\sp{-1}_{\hat
g,j}=\frac1{|\mathcal{A}|\,|\mathcal{F}|}\Cof \left(
C\sp{-1}\hat{\mathcal{A}}\right)_{j,\hat g}=
\frac1{|\mathcal{A}|}\Cof \left({\mathcal{A}}\right)_{j, g}=
{\mathcal{A}}\sp{-1}_{g,j}.
$$
Thus $$(0,\ldots,0,1)\lambda \mathcal{A}\sp{-1}=
(0,\ldots,0,1)\mathcal{A}\sp{-1}$$ where the right-hand row vector
is $g$-dimensional and the left is $\hat g$-dimensional. Bringing
these results together  establishes the theorem.

\providecommand{\bysame}{\leavevmode\hbox
to3em{\hrulefill}\thinspace}
\bibliographystyle{amsalpha}

\end{document}